\documentclass[lettersize,journal]{IEEEtran}
\usepackage{amsmath,amsfonts}
\usepackage{algorithmic}
\usepackage{algorithm}
\usepackage{array}
\usepackage[caption=false,font=normalsize,labelfont=sf,textfont=sf]{subfig}
\usepackage{textcomp}
\usepackage{stfloats}
\usepackage{url}
\usepackage{verbatim}
\usepackage{graphicx}
\usepackage{cite}
\usepackage[backref]{hyperref}
\begin{document}

\title{Parotid Gland MRI Segmentation Based on Swin-Unet and Multimodal Images}

\author{Zi'an Xu, Yin Dai, Fayu Liu, Siqi Li, Sheng Liu, Lifu Shi, Jun Fu}

\maketitle

\begin{abstract}
Background and objective: Parotid gland tumors account for approximately 2\% to 10\% of head and neck tumors. Preoperative tumor localization, differential diagnosis, and subsequent selection of appropriate treatment for parotid gland tumors are critical. However, the relative rarity of these tumors and the highly dispersed tissue types have left an unmet need for a subtle differential diagnosis of such neoplastic lesions based on preoperative radiomics. Recently, deep learning methods have developed rapidly, especially Transformer beats the traditional convolutional neural network in computer vision. Many new Transformer-based networks have been proposed for computer vision tasks. 

Methods: In this study, multicenter multimodal parotid gland MR images were collected. The Swin-Unet which was based on Transformer was used. MR images of short time inversion recovery, T1-weighted and T2-weighted modalities were combined into three-channel data to train the network. We achieved segmentation of the region of interest for parotid gland and tumor. 

Results: The Dice-Similarity Coefficient of the model on the test set was 88.63\%, Mean Pixel Accuracy was 99.31\%, Mean Intersection over Union was 83.99\%, and Hausdorff Distance was 3.04. Then a series of comparison experiments were designed in this paper to further validate the segmentation performance of the algorithm.

Conclusions: Experimental results showed that our method has good results for parotid gland and tumor segmentation. The Transformer-based network outperforms the traditional convolutional neural network in the field of medical images.

\end{abstract}

\begin{IEEEkeywords}
Transformer, deep learning, parotid gland tumor, image segmentation, multimodal image.
\end{IEEEkeywords}

\section{Introduction}
\IEEEPARstart{W}{orldwide} , head and neck tumors (cancer of the lip, oral cavity, salivary glands, pharynx, and larynx) are the eighth most common malignancies with a mortality rate of 40-60\% \cite{1-1, 1-2}. Parotid gland tumors (PGTs) are a clinically common type of head and neck tumor. It accounts for approximately 2\% to 10\% of head and neck tumors \cite{1-3}. Therefore, preoperative tumor localization, differential diagnosis, and subsequent selection of appropriate treatment for PGTs are crucial \cite{1-4, 1-5}. However, in the latest WHO classification, the relative rarity and highly dispersed tissue types of such tumors leave an unmet need for a subtle differential diagnosis of such neoplastic lesions based on preoperative radiomics \cite{1-3}.

MRI is the first choice of imaging modality for the examination of PGTs. This is because it has the ability to assess patient's anatomy without the use of ionizing radiation \cite{1-6} and has a higher resolution for imaging soft tissues, providing more information about the nature of the tumor and surrounding tissues \cite{1-7, 1-8, 1-9}. However, manual assessment of parotid status by MR images is a subjective, time-consuming, error-prone process \cite{1-10} and may produce inconsistent results \cite{1-11}. Automated image analysis methods offer the possibility of consistent, objective, and rapid diagnosis of parotid tumors \cite{1-10}. A good, automated method can significantly reduce the amount of time clinicians must spend on this task. However, segmentation of the parotid gland is particularly challenging due to its variable shape and often low contrast with surrounding structures \cite{1-12}.

Automatic segmentation models based on deep learning have been developed rapidly in recent years \cite{1-13, 1-14, 1-15}. Deep convolution models have been highly successful in biomedical image segmentation and have been introduced to the head and neck anatomy segmentation field \cite{1-16}. In recent years there are many excellent works that have attempted to implement segmentation of the parotid region using several different deep learning methods \cite{1-17, 1-18, 1-19, 1-20, 1-21, 1-22, 1-23, 1-24}. Recently, Transformer has made a breakthrough in computer vision (CV), and many new Transformer-based methods for CV tasks have been proposed \cite{1-25}. One of them is Swin-Unet, a Transformer-based segmentation network that performs well on CT images of the liver \cite{1-26}. Considering the difficulty of parotid segmentation, we implemented a Swin-Unet-based method to segment the region of interest (ROI) of parotid gland and tumor, and designed several comparative experiments to verify the effectiveness of the method in this paper.

The rest of this paper is organized as follows. Section II describes some work that is closely related to this paper. Section III describes the specific methods. Section IV presents the experiments and results. Section V provides some discussion of the experimental results. Finally, a summary of our work is presented in Section VI.

\section{Related Work}
\subsection{U-Net, the classical U-shaped structured segmentation network}
U-Net \cite{2-1} as a classical segmentation network, whose symmetric U-shaped structure has a profound influence on the subsequent network design. Until now, numerous segmentation network papers still use U-Net as a comparison experiment to evaluate the effectiveness of the network.

FCN \cite{2-2} was the first to solve the semantic segmentation problem by replacing the last fully connected layer of a traditional classification network with a convolutional layer for up-sampling. However, since the decoder of FCN is relatively simple compared to the encoder, it is slightly lacking in the details of segmentation. U-Net has made major changes in the decoder. U-Net uses multiple layers of convolution and up-sampling so that it is symmetric with the encoder structure. U-Net also uses jump connections to fuse high-resolution features from the encoder at different scales to mitigate the loss of spatial information due to down-sampling. These improvements have led to a significant improvement in the segmentation effect. And such a classical U-shaped structure also inspired the subsequent research, and various new U-shaped structure networks were proposed later.

In summary, U-Net goes beyond proposing a simple network and pioneers a new U-shaped segmented network architecture. The backbone of U-Net can be replaced by classical classification networks such as VGG \cite{2-3} or ResNet \cite{2-4} to obtain different U-Net networks. The Swin-Unet used in this paper is obtained by turning the backbone into Swin Transformer.

\subsection{Swin Transformer, a CV network based on Attention mechanism}
The Attention mechanism was first widely used in the field of Natural Language Processing (NLP). It simulates the attention model of the human brain. When people look at a thing in detail, they focus more attention on the more important parts. The Attention mechanism can then simulate this process by assigning higher weights to the more important parts of the network. The Attention mechanism has many different implementations, and one of them is the Multi-Head Attention mechanism. The Transformer \cite{2-5} based on this mechanism has been very successful on various NLP tasks.

Vision Transformer (ViT) \cite{2-6} applies Transformer to image classification tasks. Although Transformer lacks inductive bias compared with Convolutional Neural Network (CNN), when pre-trained with a large enough dataset, ViT which uses transfer learning still outperforms CNN.

Swin Transformer \cite{2-7} builds the network according to the structure of CNN, which introduces a part of the inductive bias that is missing in Transformer in terms of images. It makes the network have better results for image analysis tasks. In addition, the use of the shifted window multi-head self attention mechanism allows Swin Transformer to have a computational complexity linearly related to the number of patches. This greatly improves the computational efficiency while ensuring the information exchange between patches belonging to different windows.

\subsection{Swin-Unet, a segmentation network based on U-Net and Swin Transformer}
By replacing the backbone in U-Net with Swin Transformer and making some detail changes, the Swin-Unet \cite{1-26} is obtained. Swin-Unet is the first U-architecture segmentation network based entirely on Transformer, which consists of an encoder, a bottleneck, a decoder and jump connections. Swin-Unet uses Swin Transformer layer for feature extraction, Patch Merging layer and Patch Expanding layer for down-sampling and up-sampling, and introduces jump links for fusing encoder features in the decoder with reference to U-Net.

The paper proposing Swin-Unet uses the Synapse multi-organ CT dataset to demonstrate that the network works better for medical images. Therefore, in this paper, we designed an experiment of parotid gland segmentation based on Swin-Unet and obtained good experimental results. The specific method will be described in Section III.

\section{Methods}
\subsection{Collected Dataset}
For imaging data, the dataset collected in this study contains multicenter multimodal MR images of 148 patients with parotid gland tumors. The data set is from two centers with different imaging equipment and slice thicknesses. The dataset contains MRI sequences of three modalities, which are short time inversion recovery (STIR), T1-weighted (T1) and T2-weighted (T2), as shown in Figure Fig. 1(a), Fig. 1(b) and Fig. 1(c), respectively.

For the segmentation label, there is a parotid gland on each side of the image. The side with the tumor was selected by experienced clinicians, and the ROI of the parotid gland and tumor were outlined separately as segmentation labels, as shown in Fig. 1(d). We then innovatively split the whole parotid image into a left parotid image and a right parotid image from the middle and kept only the labeled side images in the dataset. For prediction, we also split the image into two parts to input them into the model and combined the output results. This reduces the clinician's workload of outlining the labels by half.

\begin{figure}[!t]
\centering
\includegraphics[width=2.5in]{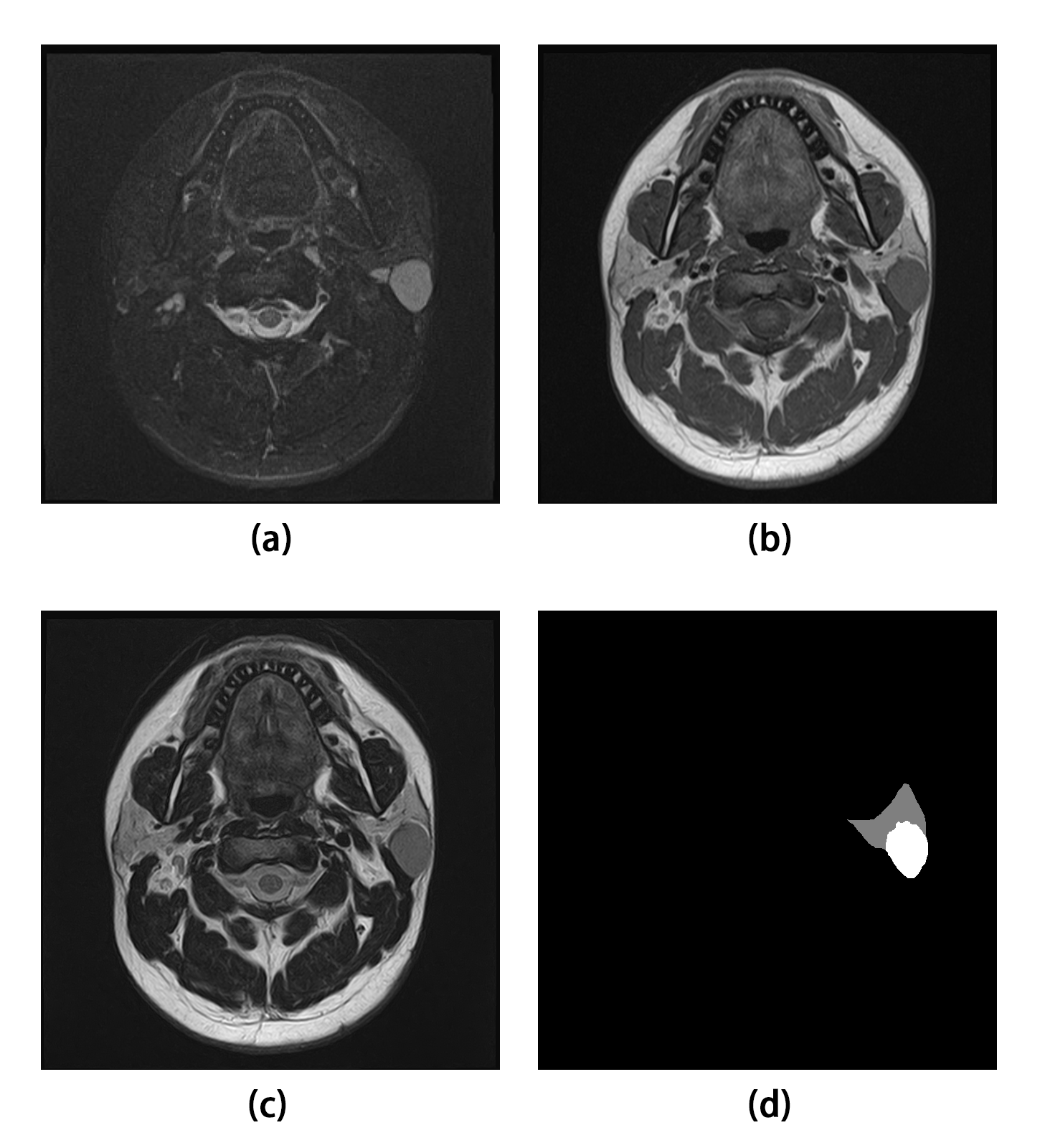}
\caption{One-layer slice from an MR image; (a) STIR image; (b) T1 image; (c) T2 image; (d) Segmentation labels, where the gray area is the parotid gland, and the white area is the tumor.}
\label{fig_1}
\end{figure}

After screening, a total of 1897 MR image slices were finally available. There were 795 image slices from the first center and 1102 image slices from the second center. In all subsequent experiments, 80\% of the used dataset was taken as the training set and the remaining 20\% as the test set.

\subsection{Method Overview}
As mentioned in Section I, the segmentation of the parotid gland is a very challenging task. And the recently proposed Transformer and its variants have shown excellent performance in natural images and other medical images. Therefore, the network used in this experiment is mainly based on the Swin-Unet introduced in Section II. The structure of the network is shown in Fig. 2.

\begin{figure}[!t]
\centering
\includegraphics[width=3.5in]{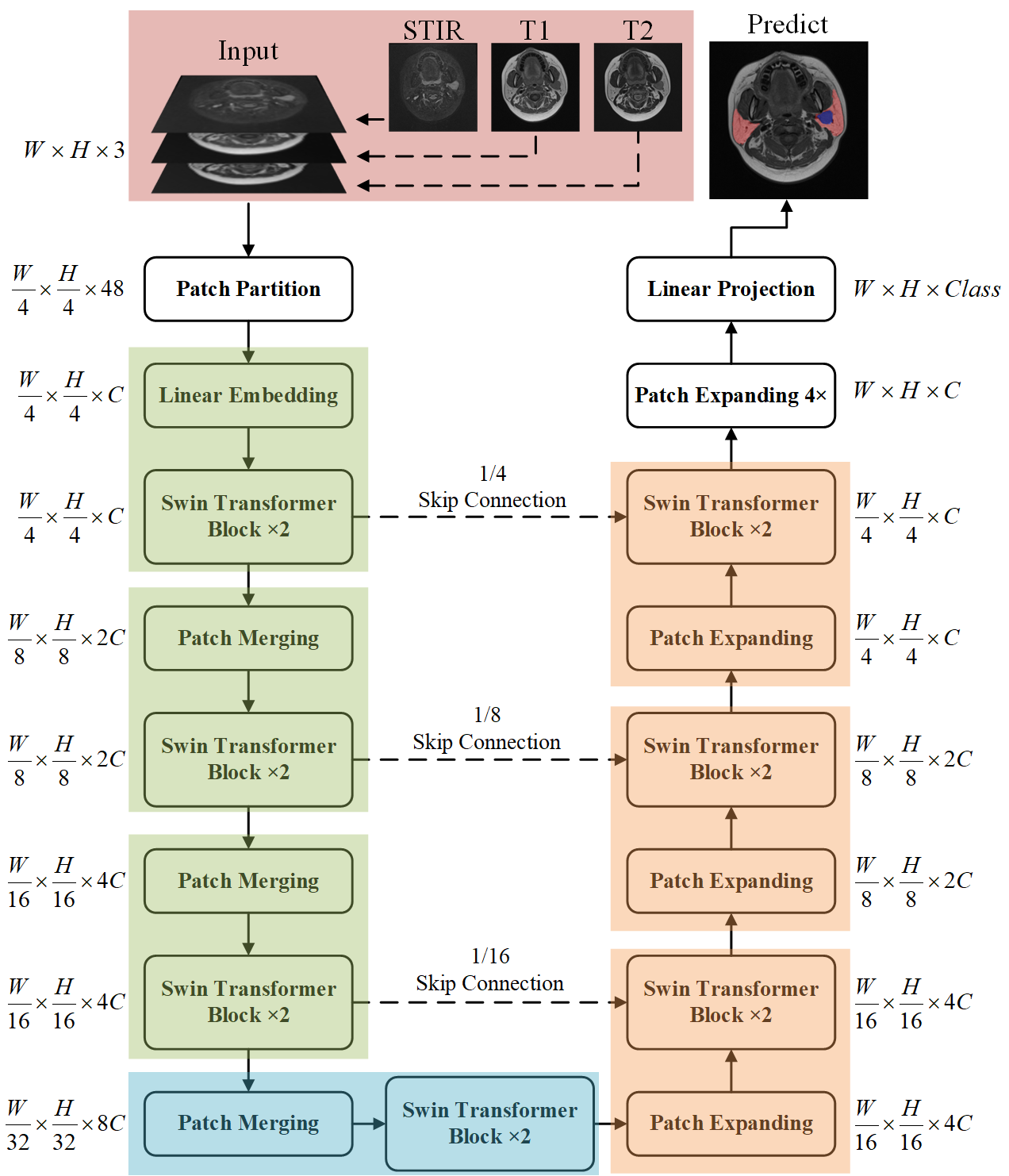}
\caption{Network structure diagram.}
\label{fig_2}
\end{figure}

In this paper, we take advantage of the multimodal STIR, T1 and T2 images from the collected MRI parotid dataset and the network structure of Swin-Unet itself, which requires a three-channel input. Each part of the network will be described in detail in turn.

\subsection{Network Inputs}
In the original paper that proposed Swin-Unet, the dataset used was the Synapse multi-organ CT. The CT images of a single layer in the dataset are single-channel grayscale images. In contrast, the input images for Swin Transformer's pre-trained model on the ImageNet dataset are three-channel. To be able to use the pre-trained model for transfer learning, the approach of the original Swin-Unet paper was to make three copies of a single-channel CT image to form a three-channel image, which was then fed into the network. Although this does not lead to an increase in information in terms of input. Instead, it increases the memory overhead. But Transformer does not have the inductive bias that CNN has. Therefore, in the case of inadequate datasets, especially for dealing with small data sets like medical images, using models pre-trained with big data for transfer learning is a necessary practice. Otherwise, the final training result of Transformer may not even be as good as traditional CNN.

In summary, using Swin Transformer's pre-trained weights on Swin-Unet's encoder for transfer learning is necessary. And the parotid data set collected in this study contains three different imaging modalities, namely STIR, T1 and T2. Therefore, we innovatively extracted the same slice from the MR image sequences obtained from the three imaging modalities and stitched them in the channel dimension to form a three-channel image. This enables both using the pre-trained model for transfer learning and inputting more image information to further improve the accuracy of the model. Although the pre-trained model originally used natural images with R, G, and B channels, which are very different in physical sense from the STIR, T1, and T2 channels we use now. However, the comparison experiments designed in Section IV prove that doing so does improve the network partitioning performance.

\subsection{Patch Partition layer and Linear Embedding layer}
Transformer was first applied in the field of NLP. To improve the speed of network computing, One-Hot encoding is often used to represent each word. However, each different word should correspond to a different One-Hot encoding will result in too large a feature dimension of the input. Therefore, it is necessary to use Embedding layers with learnable parameters for dimensionality reduction.

To use Transformer in the field of CV, it is proposed in ViT to slice an image into $16 \times 16$ independent patches and treat the image as $16 \times 16$ numbers of words input to the network. The purpose of the Patch Partition layer is to divide the image into several separate patches. In Swin-Unet used in this paper, the Patch Partition layer partitions the input of size $(H,W,3)$ into patches of size $(\frac{W}{4},\frac{H}{4},48)$. The latter also follows the steps of NLP, using Linear Embedding layers with learnable parameters for dimensionality reduction. In terms of the specific implementation, these two layers can be implemented in one step using a convolution operation with a convolution kernel size of $4 \times 4$ and a step size of $4$.

\subsection{Swin Transformer layer}
As the main feature extraction layer, Swin Transformer consists of two different Attention modules. They are the Window Multihead Self-Attention Module (W-MSA) and the Moving Window Multihead Self-Attention Module (SW-MSA), respectively. The structure of the Swin Transformer is shown in Fig. 3.

\begin{figure}[!t]
\centering
\includegraphics[width=2.5in]{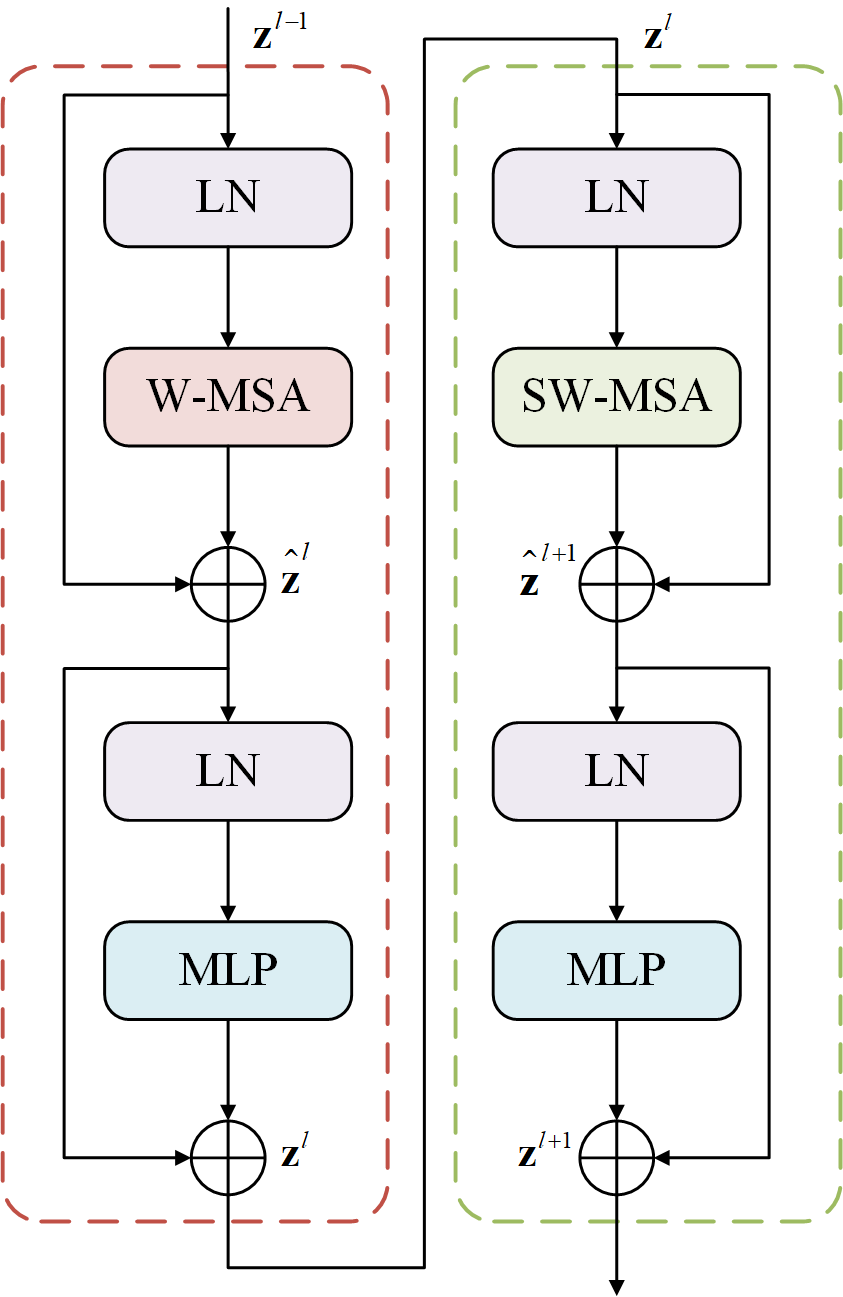}
\caption{Swin Transformer structure diagram.}
\label{fig_3}
\end{figure}

The equations for variables ${\widehat {\bf{z}}^l}$, ${{\bf{z}}^l}$, ${\widehat {\bf{z}}^{l + 1}}$ and  ${{\bf{z}}^{l + 1}}$ in the above figure are shown in equations (1), (2), (3) and (4), respectively.

\begin{equation}
\label{equ_1}
\hat{\mathbf{z}}^{l}=\mathrm{W}-\mathrm{MSA}\left(\mathrm{LN}\left(\mathbf{z}^{l-1}\right)\right)+\mathbf{z}^{l-1}
\end{equation}

\begin{equation}
\label{equ_2}
\mathbf{z}^{l}=\operatorname{MLP}\left(\operatorname{LN}\left(\hat{\mathbf{z}}^{l}\right)\right)+\hat{\mathbf{z}}^{l}
\end{equation}

\begin{equation}
\label{equ_3}
\hat{\mathbf{z}}^{l+1}=\operatorname{SW-MSA}\left(\mathrm{LN}\left(\mathbf{z}^{l}\right)\right)+\mathbf{z}^{l}
\end{equation}

\begin{equation}
\label{equ_4}
\mathbf{z}^{l+1}=\operatorname{MLP}\left(\operatorname{LN}\left(\hat{\mathbf{z}}^{l+1}\right)\right)+\hat{\mathbf{z}}^{l+1}
\end{equation}

In a Swin Transformer Block, the input data goes through the LayerNorm (LN) layer first. The LN here plays a similar role to the BatchNorm (BN), which is commonly used in the CV field. Both are designed to normalize the activation values of the output of the previous layer to avoid the gradient disappearance problem to some extent. The difference between LN and BN is the different dimensionality of the computed normalization. LN is computed in Layer dimension, while BN is computed in Batch dimension. In the NLP field, the batch size of the network is usually smaller than that in the CV field, which makes BN often less effective than LN \cite{3-1}. Therefore the LN layer is used in Transformer. The formula for LN is shown in equation (5).

\begin{equation}
\label{equ_5}
y=\frac{x-\mathrm{E}[x]}{\sqrt{\operatorname{Var}[x]+\epsilon}} * \gamma+\beta
\end{equation}

Where ${\rm{E}}[x]$ represents the mean of $x$, ${\mathop{\rm Var}\nolimits} [x]$ represents the variance of $x$, $\epsilon$ is a very small number to avoid the possibility of a zero denominator, $\gamma $ and $\beta $ are learnable parameters.

After going through the LN layer, it is entered into the W-MSA layer or SW-MSA layer. Compared to the multi-headed self-attention (MSA), W-MSA will save much computation for calculating each window separately \cite{2-7}. For an input image of size $(h,w)$, assuming that each window contains $M \times M$ patches, the computational complexity formulas for MSA and W-MSA are shown in equations (6) and (7), respectively.

\begin{equation}
\label{equ_6}
\Omega ({\rm{MSA}}) = 4hw{C^2} + 2{(hw)^2}C
\end{equation}

\begin{equation}
\label{equ_7}
\Omega ({\mathop{\rm W}\nolimits}  - MSA) = 4hw{C^2} + 2{M^2}hwC
\end{equation}

W-MSA can reduce the amount of computation, but it leads to the lack of information communication between the windows. To solve this problem, SW-MSA must be calculated in the later blocks. By moving the window down and right by half the window size and calculating W-MSA again for the moved window, it can realize the information communication between the windows. Therefore, W-MSA and SW-MSA need to appear in pairs. It is also for this reason that the number of Blocks in Swin Transformer is usually even. In Swin-Unet, the number of blocks of Swin Transformer are both 2, which contains one W-MSA block and one SW-MSA block.

After passing through the W-MSA layer or SW-MSA layer, then a BN layer, and finally a Multilayer Perceptron (MLP) for feature mapping, the final output is obtained. And to better solve the gradient vanishing problem, Swin Transformer adds residual links.

\subsection{Patch Merging layer and Patch Expanding layer}
Since the Swin Transformer layer does not result in a change in image size, the network requires a down-sampling layer and an up-sampling layer. The down-sampling and up-sampling layers in Swin-Unet are the Patch Merging layer and the Patch Expanding layer, respectively. The structure of the Patch Merging layer is shown in Fig. 4.

\begin{figure}[!t]
\centering
\includegraphics[width=3.5in]{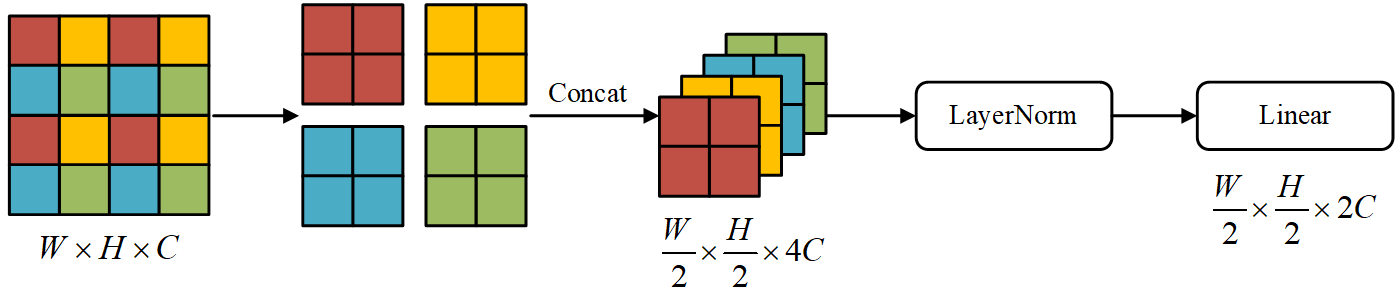}
\caption{Patch Merging structure diagram.}
\label{fig_4}
\end{figure}

Patch Merging is first done by splitting the image into regions of size $4 \times 4$, and pixels in the same position in each region are combined to form a new patch. Connecting these patches in the channel dimension, which will change the size of the input image from $\left( {W,H,C} \right)$ to $\left( {\frac{W}{2},\frac{H}{2},4C} \right)$. Now the number of channels of the image becomes four times the original. However, in U-Net, each Stage will only change the number of channels of the image to twice the original. Therefore, the number of channels needs to be halved at the end using a linear layer. To avoid the gradient vanishing problem to some extent, another LN layer is added before the linear layer, so that the whole Patch Merging layer is obtained.

The Patch Expanding layer has the opposite effect of the Patch Merging layer. It requires doubling the length and width of the image and halving the number of channels. In contrast, Patch Expanding is relatively simple. The number of channels of the input image is first doubled by a linear layer and then just rearranged. Also, to counteract gradient disappearance, an LN layer is added at the end.

\subsection{Skip Connection}
The skip connection allows features from the encoder to be fed into the decoder, thus allowing the decoder to fuse high-resolution features from the encoder at different scales to mitigate the loss of spatial information due to down-sampling. There are different ways of implementing skip connections. In FCN, a point-by-point summation method is used to sum the corresponding pixel values of the feature array directly. While, in U-Net, two feature arrays are concatenated in the channel dimension.

In Swin-Unet, skip connections are used in a slightly different way than in U-Net. After concatenating the two feature arrays in the channel dimension, an additional linear layer is required to halve the number of channels, which is required to comply with the input requirements of the next Swin Transformer layer.

\subsection{Training parameters}
In this study, Swin-Unet was used to implement the segmentation of the ROI of parotid gland and tumor in MR images. Three categories of segmentation results were set up for the network, which are background, parotid gland, and tumor. This paper investigates the medical image segmentation problem, so the classical cross-entropy function is used as the loss function. Set the batch size to 8, and the optimizer used is AdamW. The learning rate was 0.0001, and the decay rate of learning rate was 0.05. 20 epochs were trained using a transfer learning method, and the pre-training weights of the encoder part used the model parameters of the tiny version of Swin Transformer on ImageNet. The network finally achieves better training results. The test results and evaluation metrics of the model will be shown in Section IV.

\section{Experiments and results}
In this section, we first show the experimental results of the segmentation model obtained by training according to the method given in Section III. We then designed four different sets of comparison experiments for evaluating the effectiveness of the segmentation model and proving some practical conclusions.

\subsection{Experimental results of the segmentation model}
Following the training method and parameters given in Section III, after network training the segmentation model was obtained. We used four segmentation evaluation metrics to evaluate the segmentation effectiveness of the model on the test set, which are Dice-Similarity coefficient (DSC), Mean Pixel Accuracy (MPA), Mean Intersection over Union (MIoU), and Hausdorff Distance (HD).

The equations for Dice, MPA and MIoU are shown in equations (8), (9) and (10), respectively. In the formula, TP is the number of pixels correctly assigned to the category, FP is the number of pixels incorrectly assigned to the category, TN is the number of pixels correctly assigned to other categories, and FN is the number of pixels incorrectly assigned to other categories. DSC, MPA and MIoU reflect the area accuracy of the segmentation results, and higher values prove better results.

\begin{equation}
\label{equ_8}
{\rm{ Dice }} = \frac{{2TP}}{{FP + 2TP + FN}}
\end{equation}

\begin{equation}
\label{equ_9}
MPA = \frac{{TP + TN}}{{FN + TP + FP + TN}}
\end{equation}

\begin{equation}
\label{equ_10}
MIoU = \frac{{TP}}{{FN + TP + FP}}
\end{equation}

The equation of HD is shown in equation (11), where the equations of $h\left( {A,B} \right)$ and $h\left( {B,A} \right)$ are shown in equations (12) and (13), respectively.

\begin{equation}
\label{equ_11}
H(A,B) = \max (h(A,B),h(B,A))
\end{equation}

\begin{equation}
\label{equ_12}
h(A,B) = \max (a \in A)\min (b \in B)\quad \left\| {a - b} \right\|
\end{equation}

\begin{equation}
\label{equ_13}
h(B,A) = \max (b \in B)\min (a \in A)\quad \left\| {b - a} \right\|
\end{equation}

The DSC on the model test set was 88.63\%, MPA was 99.31\%, MIoU was 83.99\%, and HD was 3.04. A visual image of the partial test set segmentation results compared with the segmentation labels is shown in Fig. 5. The visualized image of partial test set segmentation results superimposed on the T1 image is shown in Fig. 6. The model has a better overall segmentation effect. However, due to the complex tissue structure of the head and neck MRI, the model still has some room for improvement in the segmentation results of slices containing smaller areas of the parotid gland.

\begin{figure}[!t]
\centering
\includegraphics[width=3.5in]{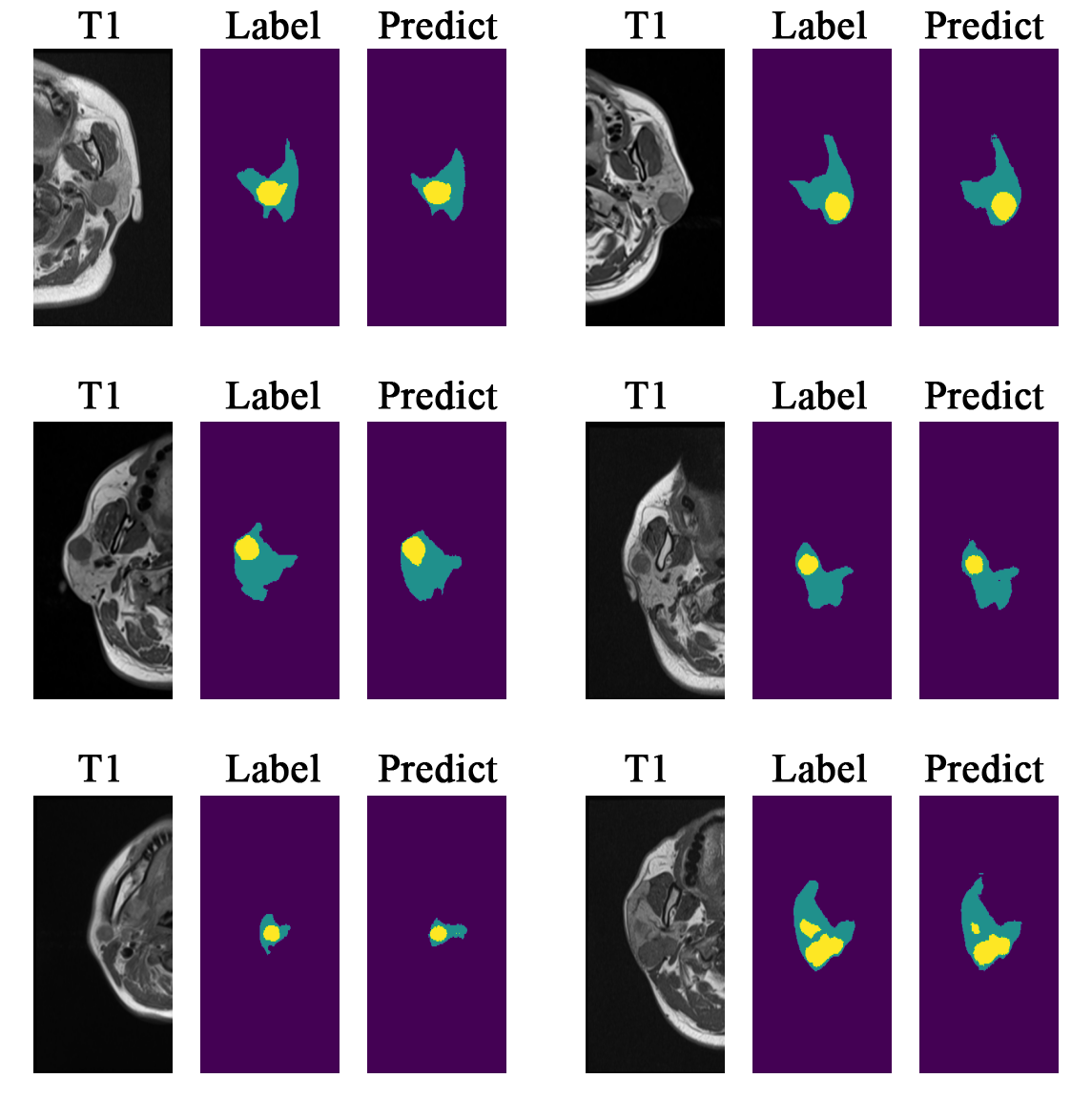}
\caption{Visual images of the segmentation results of the partial test set compared with the segmentation labels. The three images of each set from left to right are: the T1 image, the manually outlined labels and the prediction results of the segmentation model.}
\label{fig_5}
\end{figure}

\begin{figure}[!t]
\centering
\includegraphics[width=3.5in]{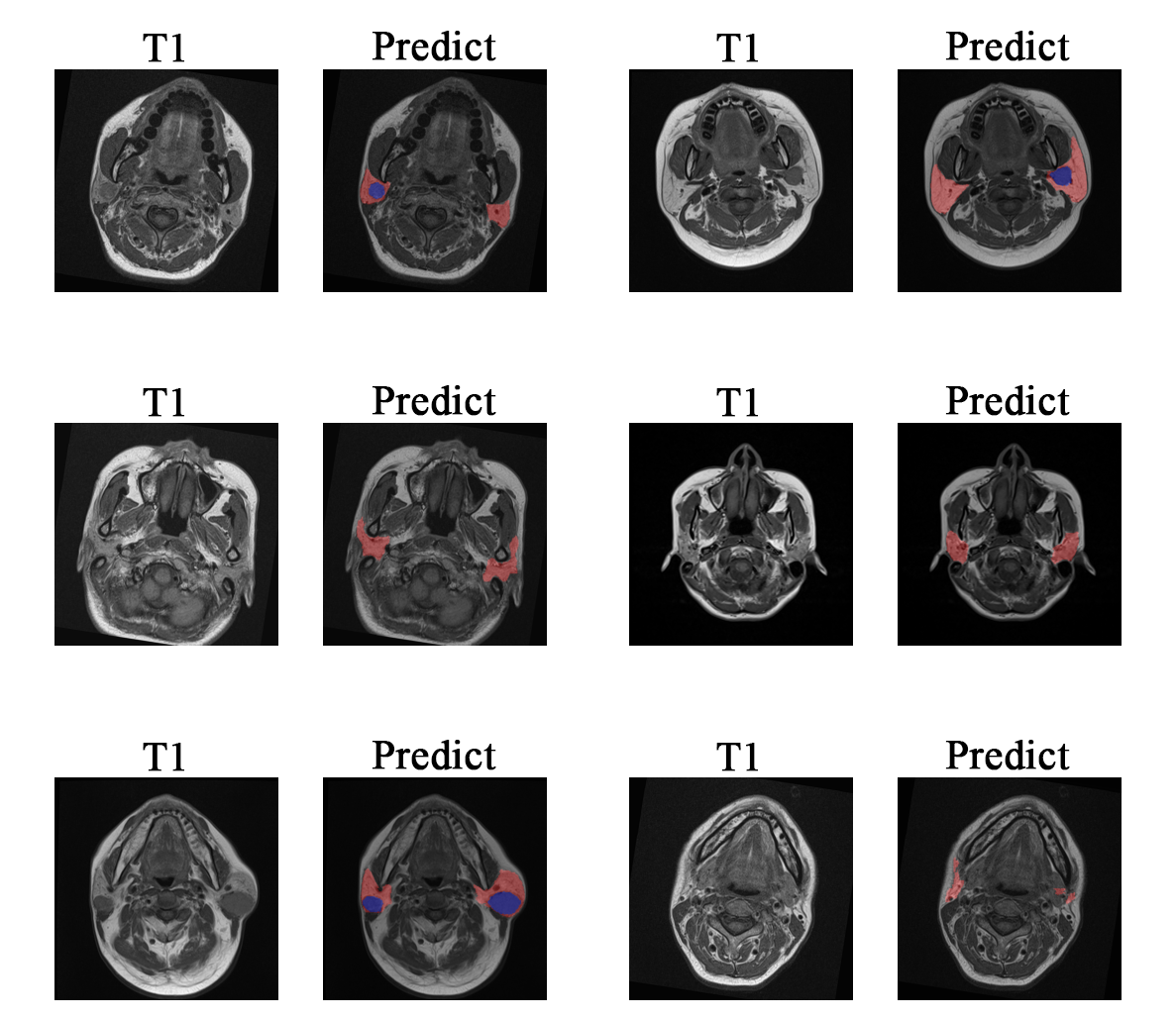}
\caption{Visualization images of partial test set segmentation results overlayed on T1. In each set of two images, the left one is the T1 image and the right one is the prediction result of the segmentation model.}
\label{fig_6}
\end{figure}

\subsection{Comparison of Swin-Unet-based and other conventional network-based model segmentation experiments}

To prove that the Swin-Unet can achieve the best results on the dataset, we trained nine classical deep learning segmentation networks including Swin-Unet using the same data, and the other eight are U-Net \cite{2-1}, UNet++ \cite{4-2}, MA-Net \cite{4-3}, LinkNet \cite{4-4}, PSPNet \cite{4-5}, PAN \cite{4-6}, DeepLabV3 \cite{4-7}, and DeepLabV3+ \cite{4-8}.

We use the same four evaluation metrics to quantify the performance of these segmentation network models. The evaluation metric values for the nine network models are shown in TABLE I. From the results of the tests, Swin-Unet was ahead of other traditional networks in all metrics.

\begin{table}
\begin{center}
\caption{Values of the four evaluation indicators \\ for the nine network models.}
\label{tab_1}
\begin{tabular}{r c c c c}
\hline
Model & DSC(\%) & MPA(\%) & MIoU(\%) & HD\\
\hline
U-Net & 87.83 & 99.25 & 83.01 & 3.09\\
UNet++ & 87.73 & 99.28 & 83.14 & 3.11\\ 
MANet & 86.52 & 99.15 & 81.34 & 3.23\\ 
LinkNet & 86.90 & 99.26 & 82.04 & 3.12\\ 
PSPNet & 84.70 & 98.99 & 79.18 & 3.38\\ 
PAN & 83.71 & 98.98 & 78.27 & 3.42\\ 
DeepLabV3 & 86.49 & 99.17 & 81.30 & 3.21\\ 
DeepLabV3+ & 87.54 & 99.19 & 82.43 & 3.10\\
\textbf{Swin-Unet} & \textbf{88.63} & \textbf{99.31} & \textbf{83.99} & \textbf{3.04}\\  
\hline
\end{tabular}
\end{center}
\end{table}

\subsection{Comparison experiment of training with STIR, T1 and T2 channels versus training with only one channel of them}
In this paper, we innovatively input the three different modalities of STIR, T1 and T2 data into Swin-Unet as a single image with three channels. To quantify the performance gains from doing so and to demonstrate that the performance gains are not the result of either modality acting alone, we designed comparative experiments with networks trained using only one type of modal images.

In this experiment, the control group used three modalities, STIR, T1, and T2, as the three-channel input network of the images. The three experimental groups took one of these three modalities in turn and made three copies as the three-channel image input network. Both experimental and control groups used transfer learning. The results of the model on the test set were evaluated and compared by continuing to use the four evaluation metrics from the previous experiment, and the results of the experiment are shown in TABLE II.

\begin{table}
\begin{center}
\caption{Results of image training using STIR, T1, and T2 channels versus using only one of the modalities.}
\label{tab_2}
\begin{tabular}{c c c c c}
\hline
Channel & DSC(\%) & MPA(\%) & MIoU(\%) & HD\\
\hline
(STIR, STIR, STIR) & 76.22 & 98.72 & 71.30 & 3.76\\
(T1, T1, T1) & 79.80 & 99.05 & 74.45 & 3.46\\ 
(T2, T2, T2) & 83.50 & 99.09 & 78.18 & 3.37\\ 
\textbf{(STIR, T1, T2)} & \textbf{88.63} & \textbf{99.31} & \textbf{83.99} & \textbf{3.04}\\  
\hline
\end{tabular}
\end{center}
\end{table}

In all four metrics, the results using the STIR, T1, and T2 channels showed a greater performance improvement than the results using images from any one of these modalities alone. This proves that using STIR, T1, and T2 modalities as a three-channel input network for images does lead to better results.

\subsection{Comparison experiments using transfer learning versus without transfer learning}
In the previous comparison experiment, it was shown that combining STIR, T1, and T2 modalities as a three-channel image input network can achieve better performance. However, Swin Transformer's pre-trained model is based on natural images trained in ImageNet with R, G and B as the three channels. After the input data three channels change from R, G and B to STIR, T1 and T2, can transfer learning still have a positive effect on the performance of the model?

To verify this question, we design a comparison experiment using transfer learning versus without transfer learning in this paper. In this experiment, the initialization weights used by the group using transfer learning were tiny versions of Swin Transformer's pre-trained model on ImageNet. The weights of the group without transfer learning are then randomly generated by the algorithm. The experimental results are shown in Fig. 7.

\begin{figure}[!t]
\centering
\includegraphics[width=3.5in]{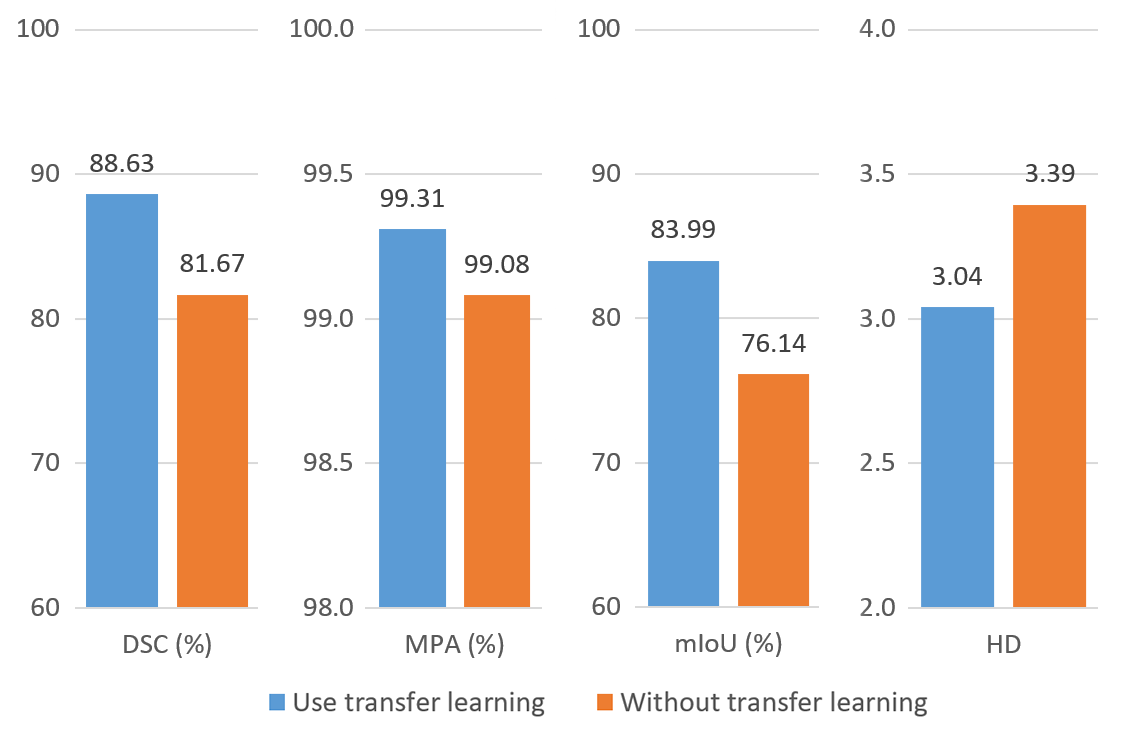}
\caption{ Results of using transfer learning and without transfer learning.}
\label{fig_7}
\end{figure}

The results show that although the three channels of input data change from R, G, and B to STIR, T1, and T2, using transfer learning still improves the performance. This may be because the Transformer-based Swin-Unet lacks some inductive bias that CNN have. For example, CNN considers image information as having spatial locality, and reduces the parameter space by sliding convolution and sharing weights. This also allows the CNN to have the generalizability of the object in terms of image location. Such a priori knowledge is what Transformer lacks. The use of transfer learning can precisely compensate for the lack of inductive bias of Transformer by pre-training with big data, and may obtain more common higher-order image features. Therefore, after the reasonable use of transfer learning, the performance of the model is better than the one without transfer learning, even better than CNN.

\subsection{Comparison experiment of the difference between different center images}
The dataset collected in this study contains two sets of images from different centers. The two sets of images differed in terms of imaging equipment and slice thickness. The practice in previous experiments was to first mix the two sets of images and then randomly select the training and test sets from them. To verify the rationality of doing so, this experiment used the above approach as a control group and designed four experimental groups with different training and test sets from each other. They are, in order, the training set using the first central image and the test set using the first central image, the training set using the first central image and the test set using the second central image, the training set using the second central image and the test set using the first central image and the training set using the second central image and the test set using the second central image.

The same four segmentation metrics were used to evaluate the model performance, and the results of the metrics for each of the four experimental groups and one control group are shown in TABLE III.

\begin{table}
\begin{center}
\caption{Results for the four control groups \\ and one experimental group.}
\label{tab_3}
\begin{tabular}{c c c c c}
\hline
Experiment group & DSC(\%) & MPA(\%) & MIoU(\%) & HD\\
\hline
train, test: random & 88.63 & 99.31 & 83.99 & 3.04\\
train: 1, test: 1 & 86.12 & 99.19 & 80.64 & 3.33\\ 
train: 1, test: 2 & 79.62 & 98.92 & 73.95 & 3.68\\ 
train: 2, test: 1 & 80.14 & 98.64 & 74.18 & 3.81\\ 
train: 2, test: 2 & 89.45 & 99.34 & 84.94 & 3.06\\ 
\hline
\end{tabular}
\end{center}
\end{table}

We find that the evaluation metrics of the model are similar to the control group when the data in the training and test sets are from the same center, and the evaluation metrics of the model and will be very different from the experimental group when the data in the training and test sets are from different centers.

Therefore, it can be concluded that data from different centers have great differences in deep learning. The training set of the network should contain data from multiple centers so that the trained model can maintain high accuracy on data from different centers.

\section{Discussion}
Deep learning networks based on Transformer lack some important inductive biases (e.g., locality and translation equivariance) compared to CNN networks, which makes training them very dependent on large-scale datasets and pre-training models \cite{5-1}. However, limited by the lack of large-scale and well-annotated datasets, the development of deep learning in the medical image domain lags behind that in the natural image domain \cite{5-2}. Especially, there are still few studies applying Transformer to the medical field \cite{5-3, 5-4, 5-5, 5-6, 5-7}. We applied the deep learning method based on Transformer to the parotid MR image segmentation task and achieved good accuracy. This further demonstrates that Transformer has outstanding results over traditional CNN networks in medical images and helps to close the development gap between deep learning in medical images and natural images.

How to apply deep learning to multimodal images has been one of the research hotspots in the field of medical imaging \cite{5-8, 5-9, 5-10, 5-11}. One of the commonly used methods is to input images of different modalities into different feature extractors and subsequently fuse the features using a fully connected layer. There are also practices of fusing different modalities of images using image fusion methods. In this paper, we propose a novel multimodal image training approach by composing three different modal images of STIR, T1, and T2 of MR images into three-channel images. Then we used ablation experiments to verify that this approach can indeed exploit information from different modalities and therefore improve the accuracy of the model. We also verify that even if it is different from the RGB channel of natural images, using the transfer learning method can still bring accuracy improvement. We consider that this simple and efficient method can be continued in future studies.

Due to the small number and the difficulty in acquiring medical images \cite{5-12, 5-13}, many medical image datasets have images from multiple centers. The data from different centers sometimes have significant differences. Several studies train different deep learning models for data from different centers \cite{5-14}. However, this may suffer from the problem of lacking the amount of data in a single center and spending a lot of training resources. In this paper, we use ablation experiments to validate the training approach for multicenter data. We found that mixing data from different centers for training enhances the generalizability of the model. And the accuracy is not significantly lower than that of the model for a single center. Therefore, we consider that the training set of the network should contain data from multiple centers so that the trained model can maintain high accuracy on data from different centers.

\section{Conclusion}
In this paper, based on Swin-Unet, three imaging modalities STIR, T1 and T2 of MRI are innovatively used as the three-channel input of the network, and a parotid segmentation model with better performance is trained. On the test set the model has a DSC of 88.63\%, MPA of 99.31\%, MIoU of 83.99\%, and HD of 3.04.

We designed four different sets of comparison experiments, which proved the following conclusions. Swin-Unet has better segmentation results than the traditional network on this dataset. Using STIR, T1, and T2 modalities as a three-channel input network is better than replicating only one of them for three channels. Transfer learning can improve model segmentation performance. The training set should contain images from all centers in the dataset.

\section*{Declaration of Competing Interest}
The authors declare that there is no conflict of interests.

\section*{Acknowledgments}
This research was funded in part by the Youth Program of National Natural Science Foundation of China grant number 61902058, in part by the Fundamental Research Funds for the Central Universities grant number N2019002, in part by the Natural Science Foundation of Liaoning Province grant number No. 2019-ZD-0751, in part by the Fundamental Research Funds for the Central Universities grant number No. JC2019025, in part by the Medical Imaging Intelligence Research grant number N2124006-3, in part by the National Natural Science Foundation of China grant number 61872075.

\bibliographystyle{IEEEtran}
\bibliography{paper}

\newpage

\begin{IEEEbiography}[{\includegraphics[width=1in,height=1.25in,clip,keepaspectratio]{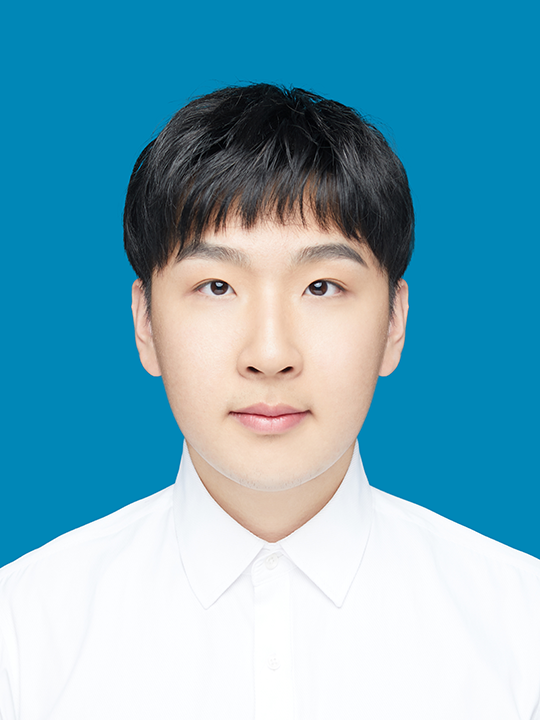}}]{Zi'an Xu}
is currently pursuing his master's degree in College of Medicine and Biological Information Engineering from Northeastern University, Shenyang, China. His research interests focus on deep learning in computer vision, classification and segmentation of medical images, and computer-aided diagnosis.
\end{IEEEbiography}

\vspace{11pt}

\begin{IEEEbiography}[{\includegraphics[width=1in,height=1.25in,clip,keepaspectratio]{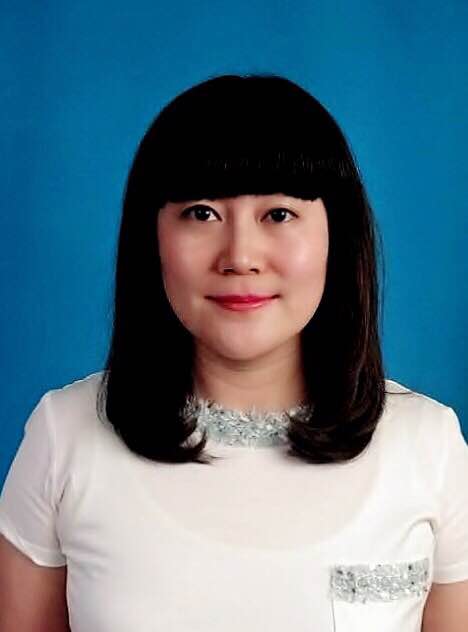}}]{Yin Dai}
received the Ph.D. degree from the Department of Computer Science, Northeastern University, China, in 2015. She is currently an associate professor in the College of Medicine and Biological Information Engineering at Northeastern University, China. Her research lies at computer-aided diagnosis and medical image processing.
\end{IEEEbiography}

\vspace{11pt}

\begin{IEEEbiography}[{\includegraphics[width=1in,height=1.25in,clip,keepaspectratio]{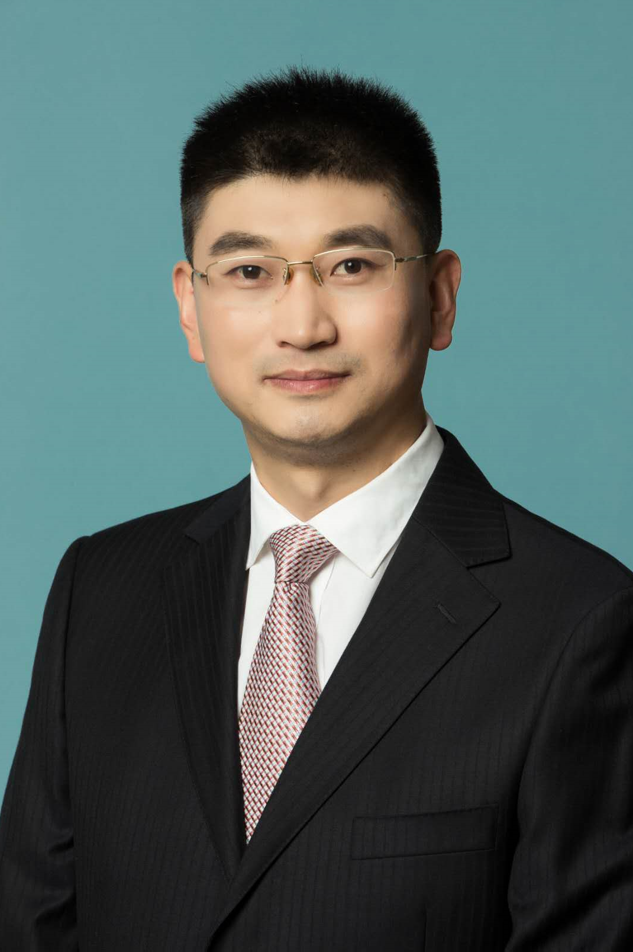}}]{Fayu Liu}
Professor, Chief Physician, Doctoral Supervisor, Director of the Department of Oral and Maxillofacial Surgery, Oral Hospital of China Medical University. He is a member of the Standing Committee of Head and Neck Specialty Committee of China Medical Education Association; a member of Skull Base Surgery Branch of China Association for the Promotion of Healthcare International Exchange; a youth member of Oral and Maxillofacial-Head and Neck Tumor Specialty Committee of Chinese Oral Medical Association; a youth member of Head and Neck Tumor Specialty Committee of Chinese Anti-Cancer Association. He has been engaged in clinical, teaching and research work in oral and maxillofacial-head and neck surgery. He has received a Research Fellowship from Sloan-Kettering Cancer Center, Rosewell Park Cancer Center and the State University of New York at Buffalo, and has led one National Natural Science Foundation, four provincial-level projects, two municipal-level projects, and participated in one national key research and development project. He has published 20 papers in foreign SCI journals and many papers in domestic core journals as the first author or corresponding author. He participated in writing the book "Repair and Reconstruction of Head and Neck Defects" published by People's Health Publishing House, and won one second prize and two third prizes of Science and Technology Progress of Liaoning Provincial Government.
\end{IEEEbiography}

\vspace{11pt}

\begin{IEEEbiography}[{\includegraphics[width=1in,height=1.25in,clip,keepaspectratio]{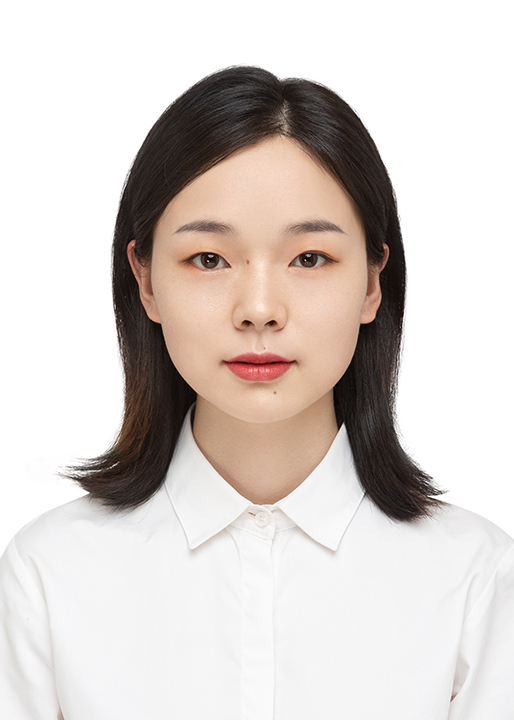}}]{Siqi Li}
graduated from the Department of Oromaxillofacial-Head and Neck Surgery, School of Stomatology of China Medical University with a master's degree in 2021.Now work in the Department of Stomatology, the children's hospital, ZheJiang University School of Mdeicine.Research interest: parotid gland tumors and radiomics.
\end{IEEEbiography}

\vspace{11pt}

\begin{IEEEbiography}[{\includegraphics[width=1in,height=1.25in,clip,keepaspectratio]{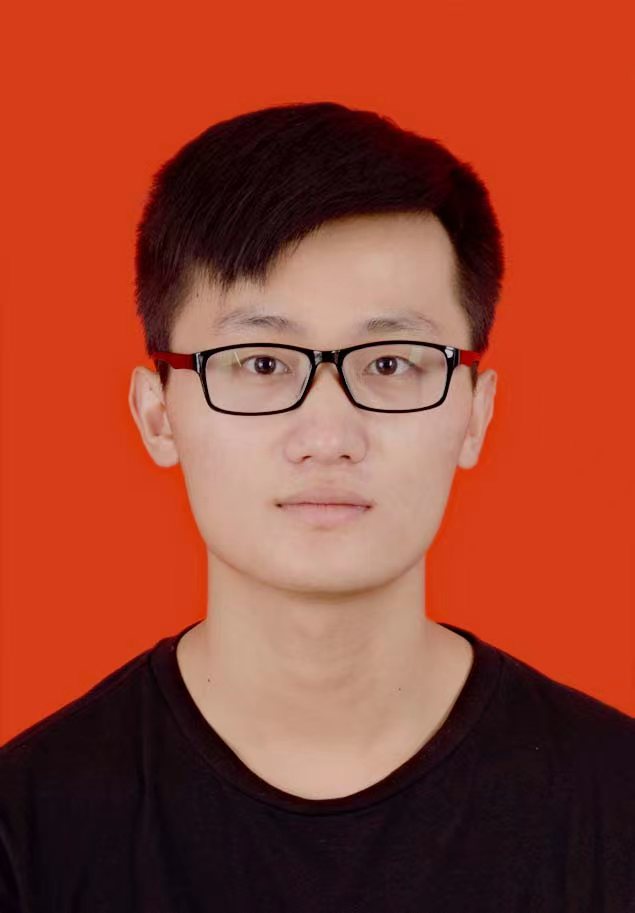}}]{Sheng Liu}
is a master student from School and Hospital of Stomatology, China Medical University. He is majoring in Oral and Maxillofacial Surgery. Currently, he is mainly engaged in research on MRI, especially the application of MRI-Based radiomics features and machine learning in predicting lymph node metastasis in the oral squamous cell carcinoma.
\end{IEEEbiography}

\vspace{11pt}

\begin{IEEEbiography}[{\includegraphics[width=1in,height=1.25in,clip,keepaspectratio]{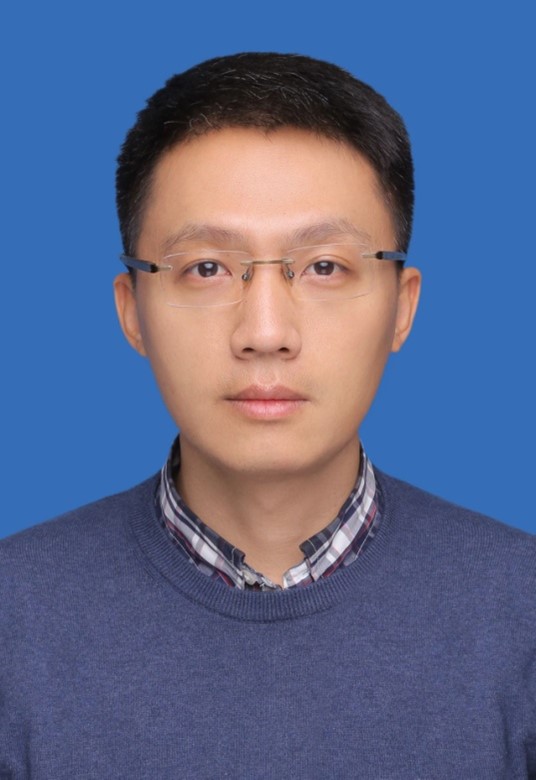}}]{Lifu Shi}
received the Master degree from the Jilin University, China, in 2013. He is mainly researching in the field of data statistics and software science . In recent years, his research mainly focuses on big data analysis and data reconstruction in the field of magnetic medical detection and treatment such as magnetic particle imaging, homogeneous magnetic sensitivity immunoassay, etc.,and also leading the development of several software systems for clinical diagnosis.
\end{IEEEbiography}

\vspace{11pt}

\begin{IEEEbiography}[{\includegraphics[width=1in,height=1.25in,clip,keepaspectratio]{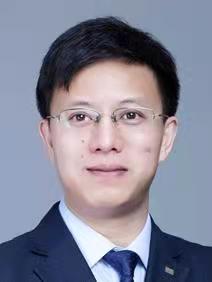}}]{Jun Fu}
received the Ph.D. degree in mechanical engineering from Concordia University, Montreal, QC, Canada, in 2009. He was a Postdoctoral Fellow/ Associate with the Department of Mechanical Engineering, Massachusetts Institute of Technology, Cambridge, MA, USA, from 2010 to 2014. He is currently a Full Professor with Northeastern University, Shenyang, China. He has authored/coauthored more than 70 publications, which appeared in journals, conference proceedings, and book chapters. His current research interests include switched systems, robust control, mathematical programming, and dynamic optimization. Dr. Fu received the 2016 Young Scientist Award from the Chinese Association of Automation. He is currently an Associate Editor of the IEEE TRANSACTIONS ON SYSTEMS, MAN, AND CYBERNETICS: SYSTEMS, IEEE TRANSACTIONS ON NEURAL NETWORKS AND LEARNING SYSTEMS, Control Engineering Practice (IFAC), and Journal of Industrial and
Management Optimization.
\end{IEEEbiography}

\vfill

\end{document}